\documentclass[12pt]{article}
\usepackage{epsfig}
\newcommand{\nc}{\newcommand}
\nc{\postscript}[2] 
{\setlength{\epsfxsize}{#2\hsize}\centerline{\epsfbox{#1}}}
\nc{\bg}{B. Grzadkowski}
\nc{\non}{\nonumber}
\nc{\hc}{\hbox {h.c.}} \nc{\re}{\hbox {Re}} 
\nc{\mev}{\hbox {MeV}} \nc{\gev}{\;\hbox {GeV}} \nc{\tev}{\;\hbox {TeV}}
\def\lsim{\mathrel{\raise.3ex\hbox{$<$\kern-.75em\lower1ex\hbox{$\sim$}}}}
\def\gsim{\mathrel{\raise.3ex\hbox{$>$\kern-.75em\lower1ex\hbox{$\sim$}}}}

\nc{\prd}[3]{{\it Phys.\ Rev.}\ {{\bf D{#1}} (#2), #3}}
\nc{\prl}[3]{{\it Phys.\ Rev.\ Lett.}\ {{\bf {#1}} (#2), #3}}
\nc{\plb}[3]{{\it Phys.\ Lett.}\ {{\bf B{#1}} (#2), #3}}
\nc{\npb}[3]{{\it Nucl.\ Phys.}\ {{\bf B{#1}} (#2), #3}}
\nc{\ptp}[3]{{\it Prog.\ Theor.\ Phys.}\ {{\bf {#1}} (#2), #3}}
\nc{\zfp}[3]{{\it Z.\ Phys.}\ {{\bf C{#1}} (#2), #3}}
\nc{\epj}[3]{{\it Eur.\ Phys.\ J.}\ {{\bf C{#1}} (#2), #3}}
\nc{\mpla}[3]{{\it Mod.\ Phys.\ Lett.}\ {{\bf A{#1}} (#2), #3}}
\nc{\rmp}[3]{{\it Rev.\ Mod.\ Phys.}\ {{\bf {#1}} (#2), #3}}
\nc{\ijmpa}[3]{{\it Int.\ J.\ of\ Mod.\ Phys.}\
               {{\bf A{#1}} (#2), #3}}
\nc{\Lsp}{\;\;\;\;\;\;\;\;\;\;}  \nc{\LLLsp}{\lspace \lspace}
\nc{\lsp}{\;\;\;\;\;\;}
\nc{\spac}{\;\;\;}
\nc{\noi}{\noindent}
\nc{\beq}{\begin{equation}}   \nc{\eeq}{\end{equation}}
\nc{\bea}{\begin{eqnarray}}   \nc{\eea}{\end{eqnarray}}
\nc{\baa}{\begin{array}}      \nc{\eaa}{\end{array}}
\nc{\bit}{\begin{itemize}}    \nc{\eit}{\end{itemize}}
\nc{\ben}{\begin{enumerate}}  \nc{\een}{\end{enumerate}}
\nc{\bce}{\begin{center}}     \nc{\ece}{\end{center}}
\def\vvis{V_{vis}\left(\frac12\right)}
\def\vhid{V_{hid}(0)}
\def\rvhid{R_{hid}V_{hid}}
\def\rhid{R_{hid}}

\def\lam{\Lambda}
\def\lamhid{\lam_{hid}}
\def\lamvis{\lam_{vis}}
\def\sig{\sigma}
\def\mpl{M_{Pl}}
\def\delt{(\delta T)_{ij}}

\def\ocal{{\cal O}}

\def\gam{{\gamma}}
\begin{document}
\pagestyle{plain}
\pagestyle{empty} \setlength{\footskip}{2.0cm}
\setlength{\oddsidemargin}{0.5cm} \setlength{\evensidemargin}{0.5cm}
\renewcommand{\thepage}{-- \arabic{page} --}

\def\mib#1{\mbox{\boldmath $#1$}}
\def\bra#1{\langle #1 |}      \def\ket#1{|#1\rangle}
\def\vev#1{\langle #1\rangle} \def\dps{\displaystyle}
\nc{\tb}{\stackrel{{\scriptscriptstyle (-)}}{t}}
\nc{\bb}{\stackrel{{\scriptscriptstyle (-)}}{b}}
\nc{\fb}{\stackrel{{\scriptscriptstyle (-)}}{f}}
\nc{\pp}{\gamma \gamma}
\nc{\pptt}{\pp \to \ttbar}
\nc{\barh}{\overline{h}}
   \def\thebibliography#1{\centerline{REFERENCES}
     \list{[\arabic{enumi}]}{\settowidth\labelwidth{[#1]}\leftmargin
     \labelwidth\advance\leftmargin\labelsep\usecounter{enumi}}
     \def\newblock{\hskip .11em plus .33em minus -.07em}\sloppy
     \clubpenalty4000\widowpenalty4000\sfcode`\.=1000\relax}\let
     \endthebibliography=\endlist
   \def\sec#1{\addtocounter{section}{1}\section*{\hspace*{-0.72cm}
     \normalsize\bf\arabic{section}.$\;$#1}\vspace*{-0.3cm}}
\vspace*{-2cm}
\begin{flushright}
$\vcenter{
\hbox{IFT-03-08}
\hbox{UCD-03-03} 
\hbox{hep-ph/0304241}
\hbox{April, 2003}
}$
\end{flushright}
\vskip 2cm
\begin{center}
{\large\bf  Bulk Scalar Stabilization of the Radion without Metric Back-Reaction in the Randall-Sundrum Model}
\end{center}

\vspace*{1cm}
\begin{center}
\renewcommand{\thefootnote} {\alph{footnote})}
{\sc Bohdan GRZADKOWSKI}\footnote{E-mail address:
\tt bohdan.grzadkowski@fuw.edu.pl}\\
Institute of Theoretical Physics, Warsaw 
University,\\
 Ho\.za 69, PL-00-681 Warsaw, POLAND\\
\vskip .6cm
{\sc John F. GUNION}\footnote{E-mail address:
\tt jfgucd@higgs.ucdavis.edu} \\
Davis Institute for High Energy Physics, 
University of California Davis, \\Davis, CA 95616-8677, USA \\
\end{center}

\vskip 1.5cm

\centerline{ABSTRACT} 
\vskip .5cm

Generalizations of the Randall-Sundrum model containing a bulk scalar
field $\Phi$ interacting with the curvature $R$ through the general
coupling $R\,f(\Phi)$ are considered.  We derive the general form of
the effective 4D potential for the spin-zero fields and
show that in the mass matrix the radion mixes with the Kaluza-Klein
modes of the bulk scalar fluctuations. We demonstrate that 
it is possible to choose a non-trivial
background form $\Phi_0(y)$ (where $y$ is the extra dimension
coordinate) for the bulk scalar field such that the exact
Randall-Sundrum metric is preserved (i.e. such that there is no
back-reaction). We compute the mass matrix for the radion and the
KK modes of the excitations of the bulk scalar relative to the
background configuration $\Phi_0(y)$ 
and find that the resulting mass matrix implies
a non-zero value for the mass of the radion (identified as the state
with the lowest eigenvalue of the scalar mass matrix).  We find that
this mass is suppressed relative to the Planck scale by the standard
warp factor needed to explain the hierarchy puzzle, implying that a
mass $\sim 1\tev$ is a natural order of magnitude for the radion mass.
The general considerations are illustrated in the case of a model
containing an $R\Phi^2$ interaction term.

\vfill



\newpage
\renewcommand{\thefootnote}{\arabic{footnote}}
\pagestyle{plain} \setcounter{footnote}{0}
\baselineskip=21.0pt plus 0.2pt minus 0.1pt
\section{Introduction}

Although the Standard Model (SM) of electroweak interactions describes
successfully almost all existing experimental data, the model
suffers from many theoretical drawbacks. One of these is the hierarchy
problem: namely, the SM cannot consistently accommodate the weak
energy scale ${\cal O}(1\tev)$ and a much higher scale such as the
Planck mass scale ${\cal O}(10^{18}\gev)$.  Therefore, it is widely
accepted that the SM is only an effective low-energy theory 
embedded in some more fundamental high-scale theory that
presumably could contain gravitational interactions.  Recently
many models that incorporate gravity have been proposed in the context
of higher dimensional space-time.
These models have received tremendous attention since they
might provide a solution to the hierarchy problem. One of the most
attractive attempts has been formulated by Randall and
Sundrum~\cite{rs} who postulated a 5D universe with two 4D surfaces
(``3-branes'') with the following action:
\beq
S=-\int d^4x\,\int_{-\frac12}^{\frac12} dy \left\{\sqrt{|g|}\left({R\over2\kappa^2}+\Lambda\right)
+\sum_{k=1,2}\sqrt{|g_k|}\lam_k \delta(y-y_k)\right\} \,,
\label{RSaction}
\eeq
where $R$ is the Ricci scalar, $\kappa^2=8\pi G_N^{(5)}$ with $G_N^{(5)}$ the Newton constant
in 5D and $\lam$, $\lam_1\equiv \lamhid$ and $\lam_2\equiv \lamvis$ 
are the cosmological constants in the bulk, on the
hidden and visible branes, respectively. In the above, 
$g_{ij}$ ($i,j=0,1,2,3,4$)
is the bulk metric and 
$(g_1)_{\mu\nu}\equiv (g_{hid})_{\mu\nu}(x)\equiv g_{\mu \nu}(x,y=y_1\equiv 0)$ and
$(g_2)_{\mu\nu}\equiv (g_{vis})_{\mu\nu}(x)\equiv g_{\mu \nu}(x,y=y_2\equiv 1/2)$ ($\mu,\nu=0,1,2,3$)
are the induced metrics on the branes.

It turns out that if the bulk and brane
cosmological constants are related by
\beq
\lam=-\frac{6m_0^2}{\kappa^2}, \spac \lamhid=-\lamvis=\frac{6m_0}{\kappa^2}
\label{RSconditions}
\eeq
and if 
periodic boundary conditions ($y\to y+1$) with identification of  $(x,y)$ and $(x,-y)$ are imposed, 
then the following metric is a solution of the 5D Einstein equations:
\beq
ds^2=e^{-2\sigma(y)}\eta_{\mu\nu}dx^\mu dx^\nu-b_0^2dy^2, 
\label{RSmetric}
\eeq
where
$\sigma(y)=m_0b_0\left[y(2\theta(y)-1)-2(y-1/2)\theta(y-1/2)\right]$;
$b_0$ is a constant parameter that is not determined by the 
equations of motion.

Within the Randall-Sundrum (RS) model all
the SM particles as well as the non-gravitational forces 
are assumed to be present on one of the 3-branes, the
``visible brane''.  Gravity lives on the visible brane, on the second
brane (the ``hidden brane'') and in the bulk.  All mass scales in the
5D theory are of the order of the Planck mass.  By placing the SM
fields on the visible brane, 
the initial 5D electroweak mass scale ${\cal O}(\mpl)$ is rescaled
by an exponential suppression factor (the ``warp factor'')
$\gamma\equiv e^{-m_0 b_0/2}$, down to the weak
scale ${\cal O}(1 \tev)$ without any severe fine
tuning. To achieve the necessary suppression, one needs $m_0 b_0 /2
\sim 37$. This is a great improvement compared to the original problem
of accommodating both the weak and the Planck scale within a single
theory. 

\section{The radion effective potential}
\label{secpotential}

A drawback of the RS model is the presence of a massless degree of
freedom called the radion. There have been several attempts (see
Refs.~\cite{radion_mass,max,casimir}) in the literature to
generate the radion mass by introducing a bulk scalar field $\Phi$
that would induce an appropriate radion potential. Here we will derive
the general form of the potential within a class of models containing
the bulk scalar interacting with gravity in the following manner:
\bea
S&=&\int d^4x\,\int_{-\frac12}^{\frac12} dy \left\{ \sqrt{|g|}\left[-{R\over2\kappa^2}-\Lambda
-\alpha R f(\Phi) + \frac12 g^{ij} \,\Phi_{, \,i}\Phi_{,\,j} -V(\Phi) \right]\non \right. \\
&&\left.- \sum_{k=1,2}\sqrt{|g_k|}[\lam_k+V_k(\Phi)] \delta(y-y_k)\right\}\, ,
\label{action}
\eea
where we have introduced the bulk potential $V(\Phi)$ and the brane
potentials $V_1(\Phi)\equiv V_{hid}(\Phi)$ and $V_2(\Phi)\equiv
V_{vis}(\Phi)$. In addition to the standard scalar kinetic-energy term,
we have allowed for a general coupling of the bulk scalar to gravity
through the $\alpha R f(\Phi)$ interaction term.
Since we would like to preserve the explanation of the hierarchy
proposed by Randall and Sundrum, we will also require that the RS
metric (\ref{RSmetric}) remain an exact solution\footnote{One can, of
  course, consider slight modifications of the RS metric that would
  also solve the hierarchy problem in a similar manner. However, in
  this paper we would like to discuss a scenario with exactly the
  same metric as in the original RS model.}  of the Einstein
equations even in the presence of the bulk scalar $\Phi$. Therefore, it is
useful to separate out in the action (\ref{action}) both the bulk
($\lam$) and brane ($\lamhid$, $\lamvis$) cosmological constants that
satisfy the RS conditions, Eq.~(\ref{RSconditions}).

In order to  identify the radion, it is sufficient 
to consider scalar excitations
of the metric  around the background RS solutions.
Hereafter, we will adopt the following parameterization 
(see Refs. \cite{Charmousis:1999rg,Csaki:2000zn})  of the metric fluctuations:
\beq
ds^2=e^{-2\sig(y)-2b(x)e^{2\sig(y)}}(\eta_{\mu\nu} + h_{\mu\nu}(x,y))
dx^\mu dx^\nu-b_0^2\left(1+2b(x)e^{2\sig(y)}\right)^2dy^2\, ,
\label{charm_metric}
\eeq
where $h_{\mu\nu}(x,y)$ and $b(x)$ are related to the
graviton\footnote{In the following we will not discuss interactions
  with gravitons as they will not influence the potential for scalar
  degrees of freedom.} and the radion, respectively.  Then from
$-\sqrt{|g|}R/(2\kappa^2)$ in the action (\ref{action}) one obtains
[after expanding in powers of $b(x)$] the kinetic term for the
radion:\footnote{Hereafter, the flat metric $\eta^{\mu\nu}$ will be
  assumed whenever repeated indices are summed.}
\beq S={6 \over
  \kappa^2 m_0}\left(e^{m_0b_0}-1\right)\int d^4x\, \frac12
\,(\partial_\mu b) \, (\partial^\mu b) + \cdots
\label{rad_kin}
\eeq
It is easy to verify that if interactions of the scalar field $\Phi$ are
switched off, then there is no potential for $b(x)$ and
consequently the radion would be massless.

The bulk scalar has been introduced here in order to generate a
non-trivial potential for the radion.  However, in general the
presence of the scalar leads  
to a non-trivial interaction potential between the
radion and the scalar in addition to the appearance of a radion
potential.  Therefore, the strategy that we will follow
here will be to determine the background scalar configuration
$\Phi(y)$ such that the RS background metric is preserved and then to
expand the action (\ref{action}) around it.  First, one has to solve
the Einstein equations together with the equation of motion for
$\Phi$.  Let us start with the Einstein equations, keeping in mind that
we would like to preserve the RS metric as a vacuum solution. We write
\beq
G_{ij}=\kappa^2\left[T_{ij}^{(RS)}+\delt(\Phi) \right]\, ,
\eeq
where $G_{ij}$ is the Einstein tensor,
$T_{ij}^{(RS)}$ denotes the RS contribution to the energy-momentum tensor and
$\delt$ contains all new contributions emerging from interactions of the scalar $\Phi$.
It is useful to first calculate all the extra (compared to the pure RS model)
contributions to the energy-momentum tensor. It is easy to show that:
\beq
\delt=T_{ij}^{(\Phi)}+2\alpha \{D_{ij}[f(\Phi)]-G_{ij}f(\Phi)\}+
\frac{1}{b_0}\sum_{k=1,2}V_k(\Phi)(g_k)_{\mu\nu}\delta^\mu_i\delta^\nu_j\delta(y-y_k)\, ,
\label{deltform}
\eeq
where
\bea
T_{ij}^{(\Phi)} &\equiv& \nabla_i \Phi \nabla_j \Phi-
g_{ij}\left[\frac12 g^{kl}\nabla_{k}\Phi\nabla_{l}\Phi -V(\Phi)\right]\\
D_{ij}[X] & \equiv & \nabla_i \nabla_j X - g_{ij} \;g^{kl} \nabla_k \nabla_l\, X \, . 
\eea
For the RS background metric we obtain:
\beq
G_{ij}=\left(
\begin{array}{ccc} 
-3 b_0^{-2} e^{-2\sig}(2\sig^{'\,2}-\sig^{''})\eta_{\mu\nu}   & &0  \\
0 & &6\sig^{'\,2}
\end{array}\right)
\label{Etens}
\eeq
and
\beq
D_{ij}[f(\Phi)]=\left(
\begin{array}{ccc} 
b_0^{-2} e^{-2\sig}(-3\sig^{'}f(\Phi)^{'}+f(\Phi)^{''})\eta_{\mu\nu}  & &0  \\
0 & &4\sig^{'}f(\Phi)^{'}\, 
\end{array}\right)\, ,
\label{Dtens}
\eeq
where here, and in what follows, the prime denotes differentiation
with respect to the 5th coordinate, $y$.

Since we demand that the RS metric be preserved  
even when the scalar is present (no back-reaction from the scalar),
we have to  require that the extra contributions to the energy momentum tensor
(calculated using the RS metric) vanish: 
\beq
\delt(\Phi)=0\, .
\label{eq12}
\eeq
Since we want to find a background solution for $\Phi$ that satisfies 
4D Lorentz invariance, we 
will assume that the solution is only a function of the 
extra dimension coordinate, $y$.
The $(\mu,\nu)$ and $(5,5)$ components of Eq.~(\ref{eq12}) read, respectively:
\bea
(\Phi^{'})^2+12\alpha(2\sig^{'\,2}-\sig^{''})f-12\alpha\sig^{'}f^{'}+4\alpha f^{''}+&&\non\\
2b_0^2\left[V(\Phi)+\frac{1}{b_0} \sum_{k=1,2}V_k(\Phi) \delta(y-y_k)\right]&=&0 \label{eq1g}\\
(\Phi^{'})^2-24\alpha\sig^{'\,2}f+16\alpha\sig^{'}f^{'}-2b_0^2V(\Phi)&=&0 \label{eq2g}
\eea
Note that since $\Phi(y)$ should be a continuous function, 
the above equations imply $\Phi^{'}(y)=0$ and $V(\Phi)=V_{vis}(\Phi)=V_{hid}(\Phi)=0$
for $\alpha=0$. Therefore, introduction of 
the extra coupling $\alpha R f(\Phi)$ is essential
in order to obtain a no-back reaction solution, $\delt(\Phi)=0$.

In addition, $\Phi$ must satisfy the following equation of motion:
\beq
-\Phi^{''}+4\sig^{'}\Phi^{'}+4\alpha (5\sig^{'\,2}-2\sig^{''}) \frac{d\,f}{d\,\Phi}+
b_0^2\left[\frac{dV}{d\Phi}+\frac{1}{b_0}\sum_{k=1,2}\frac{dV_k}{d\Phi} \delta(y-y_k)
\right]=0\, ,
\label{eqom1}
\eeq
where  the RS metric was used.\footnote{Both the 
Einstein equations (\ref{eq12}) and the equation of motion (\ref{eqom1}) 
constrain the scalar $\Phi$. However, it can 
be verified that the equations are not independent; a certain
linear combination of derivatives with respect to $y$ of the $(\mu,\nu)$ and $(5,5)$ components of
Eq.~(\ref{eq12}) is proportional to Eq.~(\ref{eqom1}). Since we have not specified the potential 
$V(\Phi)$, we can find a consistent solution both for $\Phi$ and the potential.
Details of the derivations for $f(\Phi)=3/32 \, \Phi^2$ 
will be presented in Sec.~\ref{rphi2}.}
Once the vacuum solution ($\Phi_0$) is determined, we 
expand the action, Eq.~(\ref{action}), 
adopting the parameterization of the scalar fluctuations
of the metric given in Eq.~(\ref{charm_metric}) 
and the following definition for the
$\Phi$ quantum fluctuation:
\beq
\Phi(x,y) \to \Phi_0(y) + \phi(x,y)\, .
\label{phiexcit}
\eeq 
Then, in order to determine the effective 4D potential for the scalar degrees of freedom,
we collect all non-derivative contributions 
to the $d^4x$ integrand in the action of Eq.~(\ref{action}) 
containing $b(x)$ and $\phi(x,y)$. 

In other words, we expand  the theory 
defined by the action  (\ref{action}) around the 
vacuum solutions for the metric (the RS solution) and  the bulk scalar field
$\Phi$ [the solution of Eqs.~(\ref{eq12}) and (\ref{eqom1})]
in terms of the scalar fluctuation of the metric, $b(x)$, and fluctuation
of the scalar field, $\phi(x,y)$.

First, let us calculate the Ricci scalar for the metric (\ref{charm_metric})
and collect all the terms containing derivatives with respect
the the extra component $y$:
\beq
R=\frac{20}{b_0^2} \sig^{'\;2} - \frac{8}{b_0^2} \frac{\sig^{''}}{1+2be^{2\sig}} +  \cdots
\label{R}
\eeq
where ellipses contain only $(x,y)$-derivatives
of the graviton  and $x$-derivatives of the radion. 
Since we are going to calculate the potential, 
derivatives of $b(x)$ will be dropped hereafter.
As has already been mentioned, we will not consider fluctuations of the
$\eta_{\mu\nu}$ part of the metric. Therefore,
 we will also neglect all terms containing 
$h_{\mu\nu}(x,y)$.\footnote{Adopting the traceless gauge, $h_\mu^\mu=0$, 
one can eliminate
possible mixing between $b$ or $\phi$ and $h_{\mu\nu}$. 
Consequently, graviton interactions cannot influence scalar masses. Thus,
$h_{\mu\nu}$ will be suppressed in the following.}

Using the contributions to the Ricci scalar displayed in Eq.~(\ref{R}), 
one gets the following form of the effective 4D potential from the action (\ref{action}):
\bea
V_{\rm eff}(b,\phi)&=&\int_{-\frac12}^{\frac12}dy\; e^{-4\sig -4be^{2\sig}}\times \non\\
&&\left\{
\frac12\left[\frac{1}{b_0 (1+2be^{2\sig})}(\Phi_0^{'}+\phi^{'})^2 + 
2b_0 (1+2be^{2\sig}) V(\Phi_0+\phi)\right] \right.
+ \non \\
&&\frac{2}{\kappa^2b_0} 
\left[5(1+2be^{2\sig})\sig^{'\;2} - 2\sig^{''}\right]\left[1+2\kappa^2\alpha f(\Phi_0+\phi)\right]  + \non\\
&&\left.\lam b_0(1+2be^{2\sig}) + \sum_{k=1,2}[\lam_k+V_k(\Phi_0+\phi)] \delta(y-y_k) \right\}
\label{radpot}
\eea
where $\Phi_0=\Phi_0(y)$ denotes the vacuum solution (that preserves 
the RS metric) for the scalar $\Phi$.
Note that $\Phi_0$ is determined as a solution of 
the equations of motion for the RS background
metric. As a result, 
it does not contain any dependence on the radion field $b(x)$. 
It is easy to verify that 
contributions from the pure RS model to $V_{\rm eff}(b,\phi)$
vanish when the relations (\ref{RSconditions}) are satisfied: 
a non-trivial potential requires 
an extension of the minimal RS model.

Next, it is easy to show from Eq.~(\ref{deltform}) that if $\Phi$ 
is independent of $x$ then the following identity holds:
\bea
(\delta T)_\mu^\mu(\Phi)&=& \frac{2}{b_0}\left[\frac{1}{b_0}(\Phi^{'})^2+2b_0V(\Phi)\right] +
\frac{24\alpha}{b_0^2}\left(2\sig^{'\,2}-\sig^{''}\right)f(\Phi) \non \\
&&+\frac{8\alpha}{b_0^2}\left(-3\sig^{'}f(\Phi)^{'}+f(\Phi)^{''}\right) +
\frac{4}{b_0}\sum_{k=1,2}V_k(\Phi)\delta(y-y_k)\, . 
\label{iden}
\eea
Multiplying the above equation by $\exp\{-4\sig\}(b_0/4) $, 
integrating the $(-3\sig'f+f'')$ term by parts 
and using the RS relations, Eq.~(\ref{RSconditions}), 
one obtains the following simple relation between $(\delta T)_\mu^\mu(\Phi_0)$
and the minimum of the effective potential of Eq.~(\ref{radpot}) at 
its [$b(x)=0$, $\phi(x,y)=0$] minimum:
\beq
V_{\rm eff}(0,0)=\frac{b_0}{4}\int_{-\frac12}^{\frac12}dy\; e^{-4\sig}(\delta T)_\mu^\mu(\Phi_0)\,.
\label{relation}
\eeq
Note that in Eq.~(\ref{relation}) we employ the trace from Eq.~(\ref{iden}) 
calculated for the background solution $\Phi_0$.
Since the no-back-reaction requirement, Eq.~(\ref{eq12}), 
implies $(\delta T)_\mu^\mu(\Phi_0)=0$, the relation (\ref{relation})
shows that the effective potential must vanish  at the minimum 
\beq
V_{\rm eff}(0,0)=0\, .
\label{minimum}
\eeq
It is straightforward to verify that linear terms 
in $b(x)$ and $\phi(x,y)$ disappear by virtue of
Eq.~(\ref{eq12}) and Eq.~(\ref{eqom1}), respectively.

In order to determine scalar masses one has to expand the action (\ref{action}) up to terms quadratic in
$b$ and $\phi$. First, let us define the K-K modes of the scalar fluctuations:
\beq
\phi(x,y)=\sum_n\varphi_n(x)\frac{J_n(y)}{b_0^{1/2}}\, ,
\eeq
with orthonormal functions $J_n(y)$:
\beq
\int_{-\frac12}^{\frac12}dy\; e^{-2\sig(y)}J_n(y)J_m(y)=\delta_{nm}\,.
\eeq
The resulting mass terms are the following:
\beq
\frac12
\left(
\begin{array}{cc}
r &\varphi_m
\end{array}
\right)
\left(
\begin{array}{ll} 
\phantom{(}M^2_r\phantom{)} & (M^2_{r\varphi})_n\\
(M^2_{r\varphi})_m & (M^2_{\varphi})_{mn}
\end{array}
\right)
\left(
\begin{array}{c}
r\\
\varphi_n 
\end{array}
\right)\, ,
\label{mm}
\eeq
where $r$ is the canonically normalized radion [see Eq.~(\ref{rad_kin})]:
\beq
r(x)=\left(\frac{6}{\kappa^2m_0}\right)^{1/2}e^{m_0b_0/2} b(x)
\eeq
Inputing the equation of motion (\ref{eqom1}), the elements of the mass matrix read:
\bea
M^2_r&=&\frac23\kappa^2\frac{m_0}{b_0} e^{-m_0b_0} 
\int_{-\frac12}^{\frac12} (\Phi_0^{'})^2\;dy \label{radiag}\\
(M^2_{r\varphi})_n&=&\left(\frac23\kappa^2\frac{m_0}{b_0}e^{-m_0b_0}\right)^{1/2}
\frac{1}{b_0}\int_{-\frac12}^{\frac12} e^{-2\sig(y)}J_n(y)\times\non\\
&&\left[ \Phi_0^{''}+2\sig^{'}\Phi_0^{'}+
20\alpha\sig^{'\,2}\frac{d\,f}{d\,\Phi}(\Phi_0)+b_0^2\frac{d\,V}{d\,\Phi}(\Phi_0)\right]\\
(M^2_{\varphi})_{mn}&=&\frac{1}{b_0^2}\int_{-\frac12}^{\frac12} e^{-4\sig(y)}
\Biggl\{J_n(y)J_m(y)\times\non\\
&&\left[
4\alpha (5\sig^{'\,2}-2\sig^{''}) \frac{d^2f}{d\Phi^2}+
b_0^2\left(\frac{d^2V}{d\Phi^2}+\frac{1}{b_0}\sum_{k=1,2}\frac{d^2V_k}{d\Phi^2} \delta(y-y_k)
\right)\right] + \non\\
&& J_n^{'}(y)J_m^{'}(y)\Biggr\}\,.
\eea

Before we can estimate the 
size of the elements of the mass matrix, we must discuss
first the constraint that is imposed on the model 
by the requirement of maintaining the standard
strength of classical 4D gravity. 
Adopting the metric defined by Eq.~(\ref{charm_metric}), one can
calculate the coefficient of the 4D Ricci scalar obtained for  
$g_{\mu\nu}=\eta_{\mu\nu}+h_{\mu\nu}$. In order to reproduce the standard 
result, the coefficient 
should be $M_{Pl}^2/2$. The resulting constraint is:
\beq 
M_{Pl}^2={1-\gamma^2  \over \kappa^2 m_0} + 
2 \alpha b_0 \int_{-\frac12}^{\frac12}dy\; e^{-2\sig}f(\Phi_0(y))\, ,
\label{standgrav} 
\eeq 
where $\mpl\sim 2\times 10^{18}\gev$ is the reduced Planck mass and $\gamma=\exp(-m_0 b_0/2)$.
In order to solve the hierarchy problem, one needs $m_0 b_0/2\sim 37$.
Therefore, terms of order $\gamma^2$ 
can be safely neglected in Eq.~(\ref{standgrav}). 
It is clear that the most natural 
scenario\footnote{Of course, an appropriate cancellation between contributions 
coming from parameters that differ even by many orders of magnitude is in principle also
possible. However, since we would like to preserve the solution 
of the hierarchy problem proposed by Randall and Sundrum,
we should assume that all the mass parameters in the 
fundamental 5D theory are of the same order.
Then, the only necessary fine tuning is to keep  $m_0 b_0/2\sim 37$.}  emerges
when all the mass parameters of the 5D theory are of the order of $\mpl$.
In this case, the elements of the scalar mass matrix defined by 
Eq.~(\ref{mm}) are of the following order of magnitude:
\beq
\sim\mpl^2
\left(
\begin{array}{cc} 
a \;\gamma^2 & b^{1/2}\;\gamma\\
b^{1/2}\;\gamma&1
\end{array}
\right)\, ,
\eeq
where $a$ and $b$ are calculable coefficients 
of the order of $1$. It is clear that for $m_0 b_0/2\sim 37$
the lowest scalar mass is of order:
\beq
\sim{\gamma\mpl(a-b)^{1/2}\over\sqrt2}\sim \frac{246\gev}{\sqrt{2}}\,.
\label{radmass}
\eeq
There are two essential conclusions. First, we see that
the lowest scalar mass receives the standard suppression from the warp factor 
$\gamma = \exp(-m_0b_0/2)$ that is necessary for 
the solution of the hierarchy problem. 
As a result, the mass expected in the presence of a bulk scalar $Rf(\Phi)$
interaction is of the order of the electroweak scale.
Second, it is clear that 
in order to find precise values for scalar masses 
for any particular choice of the interaction function $f(\Phi)$, one has to
take into account the mixing between the radion and the Kaluza-Klein modes of
the bulk scalar fluctuation. To fully explore the phenomenology of
the theory, it would be necessary to calculate all the entries of the mass
matrix; however, this is beyond the scope of this paper.

Finally, we close this section by reiterating the
fact that if there is no $Rf(\Phi)$ interaction, i.e.
if  $\alpha=0$,  then necessarily $\Phi_0^{'}(y)=0$ and 
$V(\Phi_0)=V_{vis}(\Phi_0)=V_{hid}(\Phi_0)=0$, which in turn 
would lead to a vanishing mass matrix.


\section{The {\boldmath $R\Phi^2$} interaction}
\label{rphi2}

In this section, we will illustrate the general discussion from 
Section \ref{secpotential}, choosing
a specific form of the interaction between the bulk scalar and gravity:
\beq
f(\Phi)=\frac{3}{32}\Phi^2\,.
\label{phi2}
\eeq 
The function $f(\Phi)$ has been normalized such that $\alpha=-1$
corresponds to a 5D conformally invariant interaction. This coupling was
discussed in various contexts by many authors in the past, see e.g.
Ref.~\cite{phi2}.

In this case, the conditions for $\delt(\Phi)=0$, Eq.~(\ref{eq12}), as written
out in Eqs.~(\ref{eq1g}) and (\ref{eq2g}), read:
\bea
-3\sig^{'}(\Phi^2)^{'}+(\Phi^2)^{''}+\frac{8}{3\alpha}(\Phi^{'})^2
+3(2\sig^{'\,2}-\sig^{''})\Phi^2 &&\non\\
+\frac{16b_0^2}{3\alpha}\left[V(\Phi)
+\frac{1}{b_0}\sum_{k=1,2}V_k(\Phi)\delta(y-y_k)\right]&=&0\,; \label{eq1}\\
\sig^{'}(\Phi^2)^{'}+\frac{2}{3\alpha}(\Phi^{'})^2-\frac{4 b_0^2}{3\alpha}V(\Phi)
-\frac{3\sig^{'\,2}}{2}\Phi^2&=&0\,. \label{eq2}
\eea
Eliminating $V(\Phi)$ from Eqs.~(\ref{eq1},\ref{eq2}),
one obtains the following equation for $\Phi$:
\beq
\sig^{'}(\Phi^2)^{'}+(\Phi^2)^{''}+\frac{16}{3\alpha}(\Phi^{'})^2-3\sig^{''}\Phi^2
+\frac{16b_0}{3\alpha}\sum_{k=1,2}V_k(\Phi)\delta(y-y_k)=0\,.
\label{eq4}
\eeq
Away from the branes, we find the solution:
\beq
\Phi_0(y)=d\left[1-ce^{-\sig(y)}\right]^{\frac{1}{\beta}}\, ,
\label{solphi}
\eeq
where $c$, $d$ are integration constants and $\beta\equiv 2+8/(3\alpha)$
is required for consistency. If $1/\beta$ is not
an integer, we must also demand
that $1-ce^{-\sig(y)}>0$ in order that $\Phi_0(y)$ be well-defined.
Recalling that $\sigma''=2m_0b_0\left[\delta(y)-\delta(y-1/2)\right]$,
and noting that $(\Phi^2)''$ will contain a term proportional to $\sigma''$,
the conditions that the coefficients of the 
$\delta$-functions in Eq.~(\ref{eq4}) vanish reduce to: 
\bea
m_0\,d^2 g(c) + \beta(\beta-2)V_{hid}(0)&=&0
\label{eqa}\\
m_0\,d^2 g(c\gamma) - \beta(\beta-2)V_{vis}\left(\frac12\right)&=&0\, ,
\label{eqb}
\eea
where $\gamma \equiv \exp(-m_0b_0/2)$, 
$g(x)\equiv (1-x)^{\frac{2}{\beta}-1}\left[x(2+3\beta)-3\beta\right]$,
$V_{hid}(0)\equiv V_{hid}[\Phi_0(0)]$ and 
$V_{vis}\left(\frac12\right)\equiv V_{vis}\left[\Phi_0\left(\frac12\right)\right]$, and we have introduced the notation $V_{vis}=V_2$ and $V_{hid}=V_1$.

Insertion of the solution $\Phi_0(y)$ into, for example, Eq.~(\ref{eq2}) 
fixes the form of the bulk potential:
\beq
V(\Phi)=\frac{3}{\beta-2}m_0^2\Phi^2\left\{\frac{4}{3\beta}\left[\left(\frac{\Phi}{d}\right)^{-\beta}-1\right]
+\frac{\beta -2}{6\beta^2}\left[\left(\frac{\Phi}{d}\right)^{-\beta}-1\right]^2-1\right\}\,.
\label{potential}
\eeq
In addition, $\Phi$ must satisfy its equation of motion as obtained
from Eq.~(\ref{eqom1}) for the form Eq.~(\ref{phi2}):
\beq
4\sig^{'}\Phi^{'}-\Phi^{''}+\frac{3\alpha}{4}(5\sig^{'\,2}-2\sig^{''})\Phi+
b_0^2\left[\frac{dV}{d\Phi}+\frac{1}{b_0}\sum_{k=1,2}\frac{dV_k(\Phi)}{d\Phi}\delta(y-y_k)
\right]=0\,.
\label{eqom}
\eeq
It is easily verified that the bulk form for $\Phi_0(y)$, Eq.~(\ref{eq4}),
 also satisfies the equation of motion in the bulk, Eq.~(\ref{eqom}).
However, cancellation of the $\delta(y)$
and $\delta(y-1/2)$ brane delta function pieces yields
matching conditions that are different from Eqs.~(\ref{eqa},\ref{eqb}).
For consistency of Eq.~(\ref{eqom}) we require:
\bea
2 m_0\, d h(c) + \beta(\beta-2)V_{hid}^{'}(0)&=&0
\label{eqc}\\
2 m_0\, d h(c\gamma) - \beta(\beta-2)V_{vis}^{'}\left(\frac12\right)&=&0\, ,
\label{eqd}
\eea
where 
$h(x)\equiv (1-x)^{\frac{1}{\beta}-1}[x(2+3\beta)-4\beta]$, 
$V_{hid}^{'}(0)\equiv d V_{hid}/d\Phi|_{\Phi=\Phi_0(0)}$ and 
$V_{vis}^{'}\left(\frac12\right)\equiv d V_{vis}/d\Phi|_{\Phi=\Phi_0\left(\frac12\right)}$.

Equations (\ref{eqa}) and (\ref{eqc}) can
be solved for the parameter $c$ in terms of $V_{hid}(0)$ and 
$V_{hid}^{'}(0)$. Two solutions are possible for $c$, specified by
\beq
c_i=f_i\left(\beta,-R_{hid}\right) \spac i=1,2\,,
\label{csol}
\eeq
where
\beq
R_{hid}\equiv \frac{V_{hid}^{'\, 2}(0)}{4 m_0 V_{hid}(0)}\,.
\label{rhiddef}
\eeq 
The functions $f_i$ denote
the two possible solutions of the quadratic equations for $c$: 
\beq
f_i(\beta,R)\equiv\frac{-B\pm\sqrt{B^2-4AC}}{2A}\, ,
\label{fdef}
\eeq
where $i=1,2$ corresponds to the $+$ and $-$ signs 
in front of the square root, respectively. 
The quantities $A,B,C$ in the above are given by: 
\bea
A&=&-(2+3\beta)\left[1+\frac{2+3\beta}{R \beta (\beta-2)}\right]\non\\
B&=&(2+3\beta)\left[1+\frac{8}{R(\beta-2)}\right] +3\beta \non \\
C&=&-\beta\left[3+\frac{16}{R(\beta-2)}\right]\,.\non
\label{ABCdef}
\eea 
Positivity of $B^2-4AC$ leads to the requirement $1-(2+3\beta)/R>0$.

Once $c$ is determined, we can compute $c\gamma$ in terms of 
$V_{vis}^{'}\left(\frac12\right)$ and 
$V_{vis}\left(\frac12\right)$ using Eqs.~(\ref{eqb}) and (\ref{eqd}).
One finds
\beq
c_i\gamma_i=f_j\left(\beta,R_{vis}\right)\spac {\rm for} \spac i=1,2\, ,
\label{cgamsol}
\eeq
where   
\beq
R_{vis}\equiv \frac{V_{vis}^{'\, 2}\left(\frac12\right)}
{4 m_0 V_{vis}\left(\frac12\right)}\,.
\label{rvisdef}
\eeq
and the appropriate branch $j=1$ or $j=2$ is determined by the need
to obtain a very small value for the warp factor $\gamma_i$,
i.e. $f_j(\beta,R_{vis})\sim 0$. The latter
is most straightforwardly achieved by requiring $C_{vis}\simeq 0$
and choosing the $j=1,+$ ($j=2,-$) 
solution for $f_j(\beta,R_{vis})$ [see Eq.~(\ref{fdef})] 
for $B_{vis}>0$ ($B_{vis}<0$), respectively. From Eq.~(\ref{ABCdef}), 
the requirement that $C_{vis}\simeq 0$ is equivalent to
\beq
\beta\simeq \ -\frac23\left(\frac{8}{R_{vis}}-3\right)\, \quad\mbox{or}\quad
R_{vis}\simeq{8\over 3(1-\frac12\beta)}\,.
\label{alphafix}
\eeq 
For $R_{vis}$ as above,\footnote{For $R_{vis}=
{8\over 3(1-\frac12\beta)}\equiv r(\beta)$
exactly, $\gam=0$. As $R_{vis}$ changes (for a fixed $R_{hid}$) 
from a value slightly larger
than this to a value slightly smaller, $\gam$ will change sign.
Whether a positive or negative shift relative to $r(\beta)$ is required
to give $\gam_i>0$ is determined
by the sign of $c_i$, which can vary.} the positivity of 
$\sqrt{B_{vis}^2-4A_{vis}C_{vis}}$ is automatic so long as
$B_{vis}$ is not extremely tiny. For $R_{vis}$ as given
in Eq.~(\ref{alphafix}), one finds $B_{vis}\simeq 3\beta/2-1$, 
so that we must use the $j=1,+$ solution for $\beta>2/3$ and the $j=2,-$
solution for $\beta<2/3$. For $\beta=2/3$, the choice becomes ambiguous.

For convenience, in what follows we denote by $A_{vis},B_{vis},C_{vis}$
the values of $A,B,C$ for $R=R_{vis}$ and by $A_{hid},B_{hid},C_{hid}$
the values of $A,B,C$ for $R=-R_{hid}$.

From Eqs.~(\ref{csol}) and (\ref{cgamsol}),
the warp factor $\gamma_i$ (and hence the distance
between the branes) for a given solution $c_i$ is given by
\beq
\gamma_i \equiv e^{-\frac{m_0 {b_0}_i}{2}}=
\frac{f_j\left(\beta,R_{vis}\right)}{f_i\left(\beta,-R_{hid}\right)}\spac {\rm for} \spac i=1,2\,,
\label{local}
\eeq
with $j=1$ for $\beta>2/3$ and $j=2$ for $\beta<2/3$, as discussed earlier.
In practice, we will require that $\gamma_i=\gamma\equiv e^{-37}$
for either choice of $i$.
Further, one can use (for example) Eq.~(\ref{eqc}) to determine $d_i$:
\beq
d_i=-\frac{\beta(\beta-2)V_{hid}^{'}(0)}{2m_0 h[f_i(\beta,-R_{hid})]}\, .
\label{diform}
\eeq
Once $d_i$, $c_i$ and $\gamma_i$ have been fixed as specified above,
Eqs.~(\ref{eqa}) and (\ref{eqb}) imply a consistency
constraint on the model parameters:
\beq
\frac{V_{hid}(0)}{V_{vis}\left(\frac12\right)}=-\frac{g[f_i(\beta,-R_{hid})]}{g[f_j(\beta,R_{vis})]}\simeq \frac{g[f_i(\beta,-R_{hid})]}{3\beta}\, ,
\label{constr1}
\eeq
where $\gam_i\sim 0$ has been used to obtain the last approximate form.
Using $\gam_i\sim 0$, Eqs.~(\ref{eqb}) and (\ref{eqd}) also simplify to
\beq
V_{vis}\left(\frac12\right)\sim -3{m_0 d_i^2\over \beta-2}\,,\quad 
V_{vis}'\left(\frac12\right)\sim -8{m_0d_i\over \beta-2}\,,
\label{smallgamlims}
\eeq
with $d_i$ as given in Eq.~(\ref{diform}) for solution branch $i$.
Note that, for a given value of $\beta$ and choice of solution branch $i$, 
fixing $V_{vis}\left(\frac12\right)$ corresponds to fixing the normalization
$d_i$ of $\Phi_0$ in terms of $m_0$ and that fixing also 
$V_{vis}^\prime\left(\frac12\right)$ then fixes both $m_0$ and $d_i$.

Finally, it is important to note
that the definition of $R_{hid}$, Eq.~(\ref{rhiddef}), 
yields the following constraint on the relative signs
of $R_{hid}$ and $V_{hid}(0)$:
\beq
{V_{hid}^{'\, 2}(0)}=4m_0 R_{hid}V_{hid}(0)>0\,.
\label{constr1a}
\eeq
Using Eqs.~(\ref{smallgamlims}) and (\ref{constr1}), the condition
Eq.~(\ref{constr1a}) can be converted to a requirement
expressed entirely in terms of $\rhid$ and $\beta$ for a given solution
branch $i$:
\beq
\frac{R_{hid}g(f_i(\beta,-\rhid)}{\beta(\beta-2)}<0\, .
\label{constr1a2}
\eeq

The conformal limit of $\alpha=-1$ ($\beta=-2/3$) requires special
treatment,\footnote{In addition to the general solutions
discussed in the main text for this case, 
there exists a special background solution,
$\Phi_0(y)\propto e^{3/2\sigma(y)}$, for which
there are no matching conditions since
the solution satisfies all the necessary equations 
everywhere, including the boundaries. For this particular
special conformally symmetric case, 
substituting the form of $\Phi_0$, as given above, into Eq.~(\ref{eq2})
leads to a vanishing bulk potential, $V(\Phi_0)=0$. Similarly, Eq.~(\ref{eq4})
gives $V_{1,2}(\Phi)=0$.  Because all the potentials are zero,
one finds $M_r^2=0$. We are only interested in cases for which
a non-zero mass is generated for the radion.}
since for this choice $A_{hid}=0$. 
In this case, $g(x)=2(1-x)^{-4}$ and $h(x)={8\over 3}(1-x)^{-5/2}$.
Eqs.~(\ref{eqa}) and (\ref{eqc}) then yield 
\beq
m_0d^2(1-c)^{-4}=-{8\over 9}V_{hid}(0)\,, \quad
m_0d(1-c)^{-5/2}=-{1\over 3}V'_{hid}(0)\,,
\label{m0dconformal}
\eeq
respectively, from which we conclude that $V_{hid}(0)<0$ and $V'_{hid}(0)<0$
are required, which also implies that [see Eqs.~(\ref{rhiddef})
and (\ref{constr1a})] $R_{hid}<0$. By combining
Eq.~(\ref{m0dconformal}) and Eq.~(\ref{rhiddef}) we find
\beq
c=1+{2\over R_{hid}}\,.
\label{cval}
\eeq
Note that $R_{hid}<-2$ is required for $0<c<1$, but that $c$
is negative for $-2<R_{hid}<0$.  There is nothing obvious
to forbid this latter choice since $(1-ce^{-\sig(y)})$ will
automatically be positive for all $y$ if $c<0$.
In an analogous spirit, 
utilizing Eqs.~(\ref{eqb}) and (\ref{eqd}), one can show that
\beq
c\gamma=1-{2\over R_{vis}}\,.
\label{cgamval}
\eeq
Combining Eqs.~(\ref{cval}) and (\ref{cgamval}), one obtains the following
result for the warp factor:
\beq
\gamma=\frac{R_{hid}}{R_{vis}}\;\frac{R_{vis}-2}{R_{hid}+2}\,.
\label{gamanom}
\eeq 
In order to have a phenomenologically acceptable small value for the
warp factor $\gamma$, either $R_{vis}\sim 2$ or $R_{hid}\sim 0$
is required. 
The remaining constraint [the analog of Eq.(\ref{constr1})] in this case reads:
\beq
\frac{V_{hid}(0)}{V_{vis}\left(\frac12\right)}=
-\left(\frac{R_{hid}}{R_{vis}}\right)^4\,.
\label{constr2}
\eeq
Combining this equation with the earlier-noted constraint that
$V_{hid}(0)<0$ 
results in the requirement that  $V_{vis}\left(\frac12\right)>0$, implying
that $R_{vis}>0$.
Combined with Eq.~(\ref{gamanom}) and
the requirement that $\gamma >0$, the only allowed choices are:
\beq
\left(0<R_{vis}<2 \;\;\ {\rm and} \;\; -2<R_{hid}<0\right) \lsp {\rm or} \lsp 
\left(R_{vis}>2 \;\; {\rm and} \;\; R_{hid}<-2\right) \,.
\eeq
(Note that Eq.~(\ref{alphafix}) does
not apply for the conformal choice of $\alpha$.) For $R_{vis}\sim 2$,
as generally needed for small $\gam$, one finds
\beq
V_{vis}\left(\frac12\right)={\frac98m_0d^2}\,,\quad
V_{hid}(0)=-\frac{9}{144}m_0d^2 R_{hid}^4\,.
\label{vvalues}
\eeq
 
\begin{figure}[p]
\bce
\includegraphics[width=8.5cm]{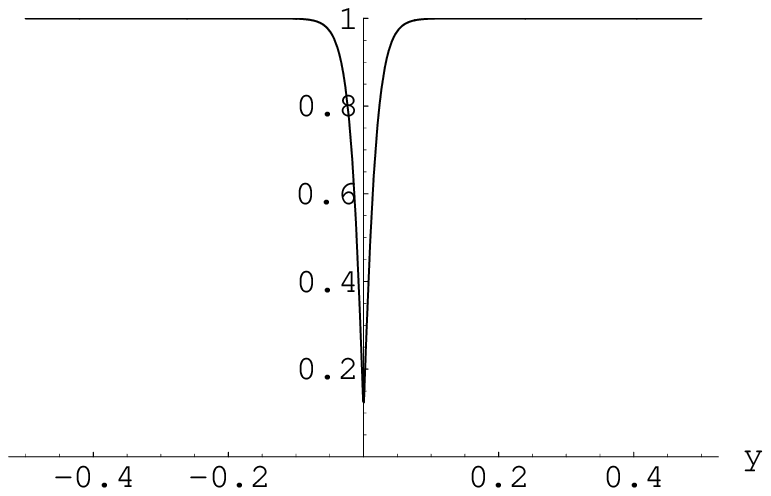}
\ece
\bce
\includegraphics[width=8.5cm]{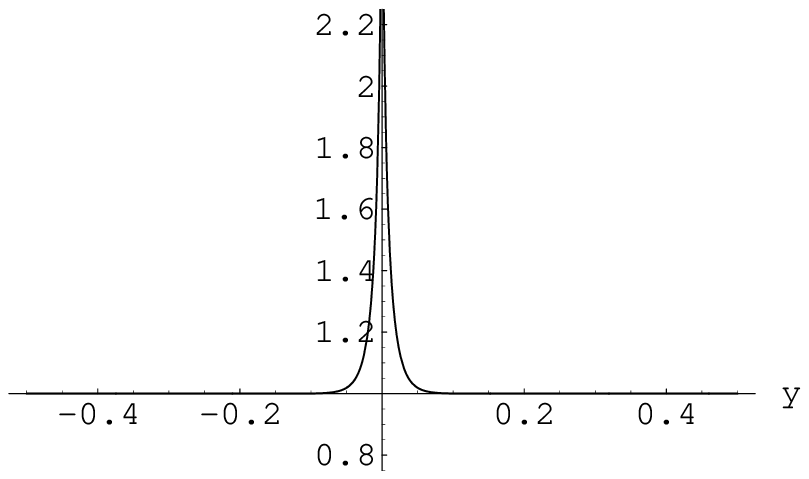}
\ece
\bce
\includegraphics[width=8.5cm]{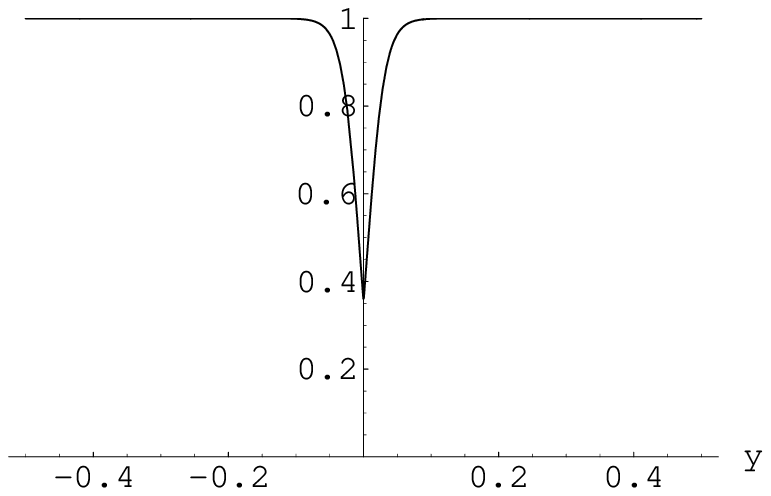}
\ece
\label{phiplots}
\caption{The normalized scalar background function 
$\Phi_0(y)/d=\left[1-ce^{-\sig(y)}\right]^{\frac{1}{\beta}}$ for 
$m_0b_0/2\sim  37$ in three cases:
(top) $R_{hid}=1$ and $\beta\sim +2/3$ (equivalent to $R_{vis}=4$);
(middle) $\beta=-2/3$ and
$R_{hid}=-4$ (implying $c=+0.5$); 
(bottom) $\beta=-2/3$ and
$R_{hid}=-1$ (implying $c=-1$). In the latter two (conformal) cases,
$m_0b_0/2=37$ requires $R_{vis}\sim 2$.}
\end{figure}

In Fig.~\ref{phiplots},  we display $\Phi_0(y)/d$
as a function of $y$ for three cases.
In all cases, we have chosen input parameters so that\footnote{We 
reemphasize that
$m_0b_0/2$ is calculable in terms of the input parameters using Eq.~(\ref{local}) or
Eq.~(\ref{gamanom}).} $m_0b_0/2\simeq 37$
as required for the warp factor $\gamma=e^{-m_0b_0/2}\sim 1\tev/\mpl$.
In the first case, we have taken $\beta\sim +2/3$, equivalent
to $R_{vis}=4$ from Eq.~(\ref{alphafix}),
and $R_{hid}=+1$. For this choice, $c\sim 0.7835$.
In the second (third) cases, we make the conformal choice of
$\beta=-2/3$, take $R_{vis}\sim 2$ (for small $\gam$) and
employ $R_{hid}=-4$ ($R_{hid}=-1$) for which $c=1/2$ ($c=-1$).
In the $\beta=2/3$ case (which is representative
of cases with $\beta>0$), 
we see that $\Phi_0(y)$ is repulsed from the hidden brane located at $y=0$.
The second (third) case is representative of a $\beta=-2/3$ case
for which $\Phi_0(y)$ is strongly peaked on (strongly repulsed from)
the hidden brane.

A useful cross-check is to adopt the 
explicit form of the solution (\ref{solphi}) to verify that
indeed $V(0,0)$ vanishes as predicted by  Eq.~(\ref{minimum}).
In order to calculate the radion potential at the minimum 
we use Eq.~(\ref{eq2}) to eliminate $V(\Phi)$ in the
general formula (\ref{radpot}). The result is the simplified expression
\bea
V_{eff}(0,0)&=&\frac{10m_0^2b_0}{\beta-2}\int_{-\frac12}^{\frac12}dy\; e^{-4\sig}(\Phi_0(y))^2
+\frac{1}{b_0}\int_{-\frac12}^{\frac12}dy\; e^{-4\sig}(\Phi_0^{'}(y))^2+\label{vradone}\\
&&\frac{8m_0}{\beta-2}\left[\Phi_0(0)^2- \Phi_0\left(\frac12\right)^2e^{-2m_0b_0}\right]
+V_{hid}(0)+V_{vis}\left(\frac12\right)e^{-2m_0b_0}\non
\eea
Then, inserting the solution Eq.~(\ref{solphi}) into Eq.~(\ref{vradone}),
one obtains:
\bea
&&\hspace*{-.7in}\beta(\beta-2)V_{eff}(0,0)\non\\
&=&
\left[m_0\,d^2(1-c)^{\frac{2}{\beta}-1}\left[c(2+3\beta)-3\beta\right]+\beta(\beta-2)V_{hid}(0)\right]\non \\
&&-\gamma^4\left[m_0\,d^2(1-c\gamma)^{\frac{2}{\beta}-1}\left[c\gamma(2+3\beta)-3\beta\right]-
\beta(\beta-2)V_{vis}(1/2)\right]\non\\
&=&\left[m_0d^2g(c)+\beta(\beta-2)V_{hid}(0)\right]
-\gamma^4\left[m_0d^2g(c\gamma)-\beta(\beta-2)V_{vis}(1/2)\right]\,.\non\\
\label{vradtwo}
\eea
The two bracketed expressions on
the right hand side of the above equation vanish by virtue of the fact that
they correspond precisely to the expressions appearing in the matching
conditions that emerge from the requirement of $(\delta T)_{\mu\nu}=0$, 
Eqs.~(\ref{eqa},\ref{eqb}).

Finally, if we insert the solution Eq.~(\ref{solphi}) into the formula 
for the radion mass, Eq.~(\ref{radiag}), one 
obtains [after employing
Eqs.~(\ref{eqa}) and (\ref{eqb})] the following result for $M_r^2$:
\beq
M_r^2=\frac23\kappa^2m_0\gamma^2\left[
-V_{hid}(0)\frac{c(2-\beta)+\beta}{c(2+3\beta)-3\beta}
-V_{vis}\left(\frac12\right)\frac{c\gamma(2-\beta)+\beta}{c\gamma(2+3\beta)-3\beta}\right]\, ,
\label{radmassphi2}
\eeq
where $c$ and $c\gamma$ are given by Eqs.~(\ref{csol}) and (\ref{cgamsol})
or, in the conformal case, 
Eqs.~(\ref{cval}) and (\ref{cgamval}).
Note the
presence of the warp factor $\gamma$ that reduces the radion mass from
the typical 5D scale [a natural choice for which
would be ${\cal O}\left(\mpl\right)$
if $V_{hid}(0)$ and $V_{vis}\left(\frac12\right)$ are $\ocal \left(\mpl^4\right)$] down to the electroweak scale. Of course, 
Eq.~(\ref{radiag}) guarantees that $M_r^2\geq0$
for any given form of $f(\Phi)$ and 
any  background solution $\Phi_0(y)$ 
that is {\it real} and fully consistent. The conditions
for reality and consistency are: (a) the constraint
Eq.~(\ref{constr1}) or, in the conformal case,
Eq.~(\ref{constr2}) is satisfied; (b) 
the positivity condition Eq.~(\ref{constr1a}) is satisfied;
and (c) $1-c>0$ when $1/\beta$ is
not an integer. However, substantial variation of $M_r^2$ is possible.
In particular, $M_r^2=0$ at special points when considered
as a function of $R_{hid}$ at fixed $\beta$.
A final form for $M_r^2$ can be derived in the $\gam\ll 1$ limit
by using Eqs.~(\ref{constr1}) and (\ref{smallgamlims}) 
in Eq.~(\ref{radmassphi2}):
\bea
M_r^2&=&{2\over 9}\kappa^2 m_0\gam^2 \vvis \left[1-
{g(c)\over\beta} {c(2-\beta)+\beta\over c(2+3\beta)-3\beta}\right]\non\\
&=& {2\over 3}(\kappa m_0 d \gam)^2{1\over \beta-2}\left[
(1-c)^{{2\over\beta}-1} \left(1-c+c{2\over \beta}\right)-1\right]\,.
\label{mrsq4}
\eea
This shows the fundamental importance of the scale of $\vvis$
in determining $M_r^2$. Of course, one should continue to keep
in mind the relation Eq.~(\ref{smallgamlims}) between $\vvis$
and $m_0d^2$ as well as the relation Eq.~(\ref{constr1}) between
$\vvis$ and $\vhid$.

\begin{figure}[h]
\bce
\includegraphics[width=10cm]{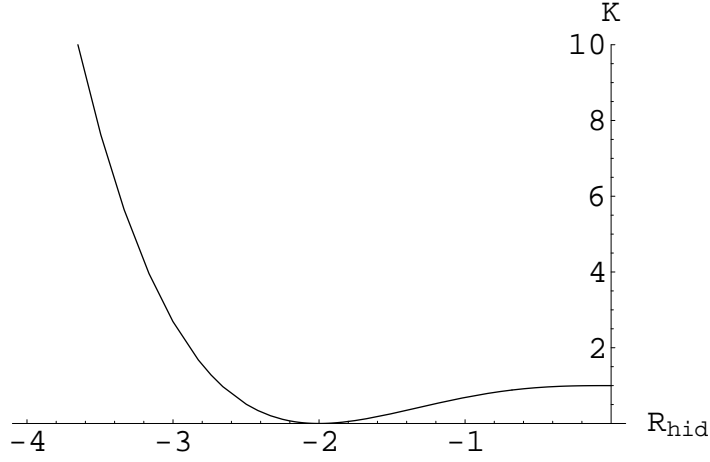}
\ece
\caption{The function $K(\rhid)$ is plotted as a function
of $R_{hid}$ in the (allowed) $\rhid<0$ region.
}
\label{rhidplot}
\end{figure}

To illustrate how $M_r^2$ depends upon
the parameters and how other constraints enter, 
consider first the conformal case with $\beta=-2/3$.
For $R_{vis}\sim 2$ (so that $\gam\ll 1$) and using
Eqs.~(\ref{cval}) and (\ref{constr2}), one  
can rewrite Eq.~(\ref{mrsq4}) in terms of $\vvis$
(which, as noted earlier, fixes the value of $d$ in terms of $m_0$).
The result is:
\beq
M_r^2=\frac29\kappa^2m_0\gam^2\vvis K(\rhid)=
\frac14(\gam \kappa m_0 d)^2K(R_{hid})\,,
\eeq
where
\beq
K(R_{hid})\equiv {3\over 16}R_{hid}^4+{1\over 2}\rhid^3+1
>0\,.
\eeq
In Fig.~\ref{rhidplot}, we plot $K(\rhid)$ as a function of $\rhid$.
We observe that
$M_r^2>0$ everywhere except at $R_{hid}=-2$ (a point where $c=0$
is required by Eq.~(\ref{cval}) and $\Phi_0(y)$ becomes trivial).
Our two earlier $\beta=-2/3$ plots of the wave function
thus correspond to choices for which $M_r^2>0$.

For fixed $\beta\neq -2/3$, and $\gam\ll 1$, $M_r^2$ is approximately
a function of $R_{hid}$ only, where $\rhid$ is to be
restricted to those values
such that a given solution $c_1$ or $c_2$ satisfies the
other consistency constraints. 
We explore the behavior of $M_r^2$ as follows.
First, recall that $V_{vis}\left(\frac12\right)$, $V_{hid}(0)$,
$R_{vis}$, $R_{hid}$ and $\beta$ are the input model parameters.
At fixed $\beta$, in order to obtain
$\gam\ll 1$, as desired on phenomenological grounds, we adjust 
$R_{vis}$ according to Eq.~(\ref{alphafix}).
Then, for a chosen $R_{hid}$  we calculate $c_i$ using Eq.(\ref{csol}). 
In fact, there are the two
solution branches, $c_{1,2}(R_{hid})=f_{1,2}(\beta,-R_{hid})$. 
Positivity of $B_{hid}^2-4A_{hid}C_{hid}$, as determined
by the sign of $1-(2+3\beta)/(-R_{hid})$, requires
$R_{hid}<{\rm Min}\left[-(2+3\beta),0\right]$ or 
$R_{hid}>{\rm Max}[0,-(2+3\beta)]$
(the limit cross over taking place at the conformal point
of $\beta=-2/3$).
Once $c_1$ ($c_2$) is chosen, if $1/\beta$ is not an integer we check 
to see if $1-c_1>0$ ($1-c_2>0$),
as required for a well-defined $\Phi_0(y)=d(1-ce^{-2\sigma(y)})^{1/\beta}$.
Finally, for any values of $\beta$ and $R_{hid}$, 
and given choice of branch $i$, we check the positivity requirement
of Eq.~(\ref{constr1a}).
To compute $M_r^2$, we adopt Eq.~(\ref{mrsq4}), which 
takes into account the consistency constraint (\ref{constr1}) such that 
$\vhid$ is expressed in terms of $V_{vis}\left(\frac12\right)$.
Eq.~(\ref{mrsq4}) makes it clear that,
for a given value of $\beta$, $M_r^2$ is proportional 
to $\vvis$ [which fixes $m_0 d_i^2$ through Eq.~(\ref{smallgamlims})] 
and depends non-trivially on $R_{hid}$, the other 
parameters being held fixed. 
Remarkably, one finds $(M_r^2)_i>0$ (as expected) so long
as: (a) $R_{hid}$ is such that $1+(2+3\beta)/R_{hid}>0$
(so that $c_{1,2}$ are real); (b) $1-c_i>0$ when 
$1/\beta$ is not an integer; and (c) 
the positivity condition, Eq.~(\ref{constr1a}), 
which we abbreviate as $\rvhid>0$, is satisfied.
There are many different
cases. Here we simply describe a couple of illustrative possibilities.

\begin{figure}[h]
\bce
\includegraphics[width=10cm]{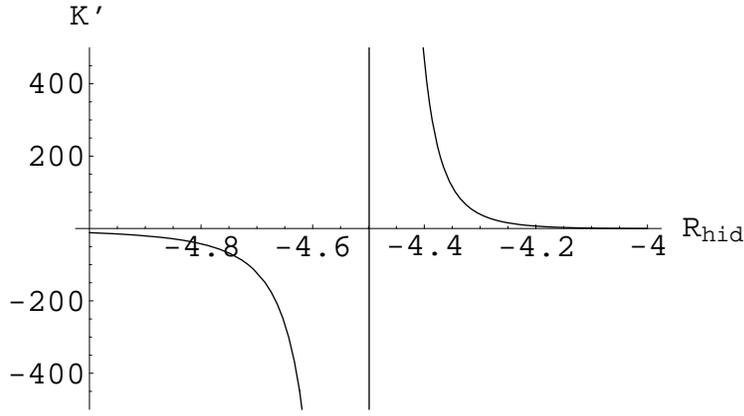}
\ece
\caption{For $\beta=2/3$, we plot
$[K']_2\equiv 
[M_r^2/\{{2\over 3}(\kappa m_0d\gam)^2\}]_2$, see Eq.~(\ref{mrsq4}),
as a function of $R_{hid}$ in the $R_{hid}\leq -(2+3\beta)=-4$ region.
Both $1-c_2$ and $(\rvhid)_2$ display the same sign changes
and singular behavior 
as $[M_r^2]_2$. Only the $[K']_2>0$ region corresponds
to a solution consistent with all constraints.
}
\label{beta2o3}
\end{figure}

Consider first two choices of $\beta$ such that $2+3\beta>0$.
\bit
\item
For $\beta=2/3$, any value of  $R_{hid}>0$ gives $1-c_{1,2}>0$ 
and $(\rvhid)_{1,2}>0$
and substantial values for the corresponding $(M_r^2)_1$ and $(M_r^2)_2$. 
For $R_{hid}<-(2+3\beta)=-4$, any value of $R_{hid}$ gives $1-c_1>0$
and substantial $(M_r^2)_1$. In all these cases, $M_r^2$
is a relatively smooth function of $\rhid$. 

However, in the $\rhid\leq -4$ region,
$1-c_2$ and $(\rvhid)_2$ are both only positive for
$-4.5\lsim R_{hid}\leq -4$ and $(M_r^2)_2$ varies rapidly,
as illustrated in Fig.~\ref{beta2o3}.
This case illustrates the extreme sensitivity that $M_r^2$
can have to $\rhid$ and shows that very large and very small
values of $M_r^2$ are quite possible.

\item
For $\beta=1/2$, since $1/\beta$ is an integer, the sign of $1-c$ is unconstrained.
One finds $(\rvhid)_{1,2}>0$ 
for all $\rhid>0$ and $(M_r^2)_{1,2}>0$ and behaves smoothly.
For all $\rhid<-7/2$, $(\rvhid)_1<0$ and this solution
branch fails the positivity constraint.
In the small region $-4.65\lsim\rhid<-7/2$, $(\rvhid)_2>0$ and $(M_r^2)_2>0$,
blowing up at $\rhid\sim -4.65$.

\eit
We also briefly describe one interesting $2+3\beta<0$ case.
\bit
\item
For $\beta=-1$, again the $1-c>0$ constraint
is not necessary. For all
$R_{hid}\geq -(2+3\beta)=1$, both $(\rvhid)_1<0$ and $(\rvhid)_2<0$
so that the $c_1$ and $c_2$ solutions both
fail the positivity constraint.   
In contrast, for all $R_{hid}<0$ one finds $(\rvhid)_1>0$ and, of course,
$(M_r^2)_1>0$. However, $(\rvhid)_2>0$ only for $-0.32\lsim\rhid<0$
and $\rhid\lsim -1.78$ --- in these two ranges $(M_r^2)_2>0$ 
aside from a zero at the top end, $\rhid=-1.78$, of the 2nd range.

\eit

Overall it is clear that there is a large range of possible models
that satisfy all the constraints necessary for exact Randall-Sundrum
metric with positive radion mass-squared. (Some particular choices
are somewhat more fine-tuned than others.) We have not understood
how to choose between the various models; possibly, the conformal
models with $\beta=-2/3$ should be regarded as the more attractive.  

\section{Conclusions}
\label{concl}

We have considered a class of generalizations of the Randall-Sundrum
model containing a bulk scalar field $\Phi$.  We demonstrated that
no-back reaction from the scalar on the Randall-Sundrum metric
solution {\it requires} the existence of an extra interaction between
gravity and the scalar. Here, we considered 
the coupling $R\,f(\Phi)$.
A general form of the potential for the fluctuation of the
compactification volume (the radion) and the Kaluza-Klein modes of the
excitation of the bulk scalar was derived and the mass matrix was
determined. In order to obtain the values of
scalar masses, one has to take into account the mixing between the radion
and the Kaluza-Klein modes of the fluctuation of the bulk scalar.
We demonstrated that a non-zero mass for the lowest eigenstate
(which we identify with the physical radion) can be 
generated using a choice of the background bulk field, $\Phi_0(y)$,
that preserves the RS metric (no back-reaction). We found that
the radion mass receives the same suppression from the warp
factor that is necessary to explain the hierarchy puzzle. Thus,
$\sim 1\tev$ is a natural order of magnitude for the radion mass.  Finally,
we illustrated the general scenario for the case of $f(\Phi)\propto
\Phi^2$, for which the scalar background solution that preserves the
Randall-Sundrum metric was explicitly found.

Since the mass-squared matrix for the radion and KK bulk scalar excitations
is non-diagonal, it is clear that the introduction of  Higgs-radion mixing
on the visible brane through a term in the Lagrangian of the form 
$\xi\sqrt{|g_{vis}|}R(g_{vis})\widehat H^\dagger \widehat H$,
as considered for example in \cite{Dominici:2002jv}, would result
in a complicated situation where the Higgs field, the radion
and the KK excitations of the bulk scalar would all mix. 
A phenomenological study of the magnitudes of the various mixings,
as a function of the available parameters,
would be required to understand the extent to which the phenomenology
of the physical Higgs boson eigenstate can be modified. Such a
study is beyond the scope of this paper.

\vspace*{0.6cm}
\centerline{ACKNOWLEDGMENTS}

\vspace*{1cm}

B.G. thanks Steve Carlip, Zygmunt Lalak, Krzysztof Meissner, Jacek Pawe\l czyk,
Manuel Toharia,  Jos\'e Wudka and Sachindeo Vaidya for useful discussions. 
B.G. is especially grateful to Jacek Tafel for reading an early version of the manuscript
and for his valuable comments. B.G. is also
indebted to the Yukawa Institute  and the Davis Institute for High
Energy Physics for their warm hospitality extended to him
while this work was being performed.
B.G. is supported in part by the State Committee for Scientific
Research under grant 5~P03B~121~20 (Poland). J.F.G. is supported
by the U.S. Department of Energy and by the Davis Institute for High
Energy Physics. This work was also supported by a joint Warsaw-Davis
collaboration grant from the National Science Foundation.

\vspace*{1cm}



\begin{thebibliography}{99}
%
\bibitem{rs}
L.~Randall and R.~Sundrum,
Phys.\ Rev.\ Lett.\  {\bf 83}, 3370 (1999)
[arXiv:hep-ph/9905221];
%
Phys.\ Rev.\ Lett.\  {\bf 83}, 4690 (1999)
[arXiv:hep-th/9906064].
%
\bibitem{radion_mass}
W.~D.~Goldberger and M.~B.~Wise,
Phys.\ Rev.\ Lett.\  {\bf 83}, 4922 (1999)
[arXiv:hep-ph/9907447];
%
T.~Tanaka and X.~Montes,
Nucl.\ Phys.\ B {\bf 582}, 259 (2000)
[arXiv:hep-th/0001092];
%
U.~Mahanta,
arXiv:hep-ph/0006350;
%
J.~M.~Cline and H.~Firouzjahi,
Phys.\ Rev.\ D {\bf 64}, 023505 (2001)
[arXiv:hep-ph/0005235];
%
B.~Grinstein, D.~R.~Nolte and W.~Skiba,
Phys.\ Rev.\ D {\bf 63}, 105016 (2001)
[arXiv:hep-th/0012202].
%
\bibitem{max}
P.~Kanti, K.~A.~Olive and M.~Pospelov,
Phys.\ Lett.\ B {\bf 481}, 386 (2000)
[arXiv:hep-ph/0002229];
%
P.~Kanti, K.~A.~Olive and M.~Pospelov,
Phys.\ Lett.\ B {\bf 538}, 146 (2002)
[arXiv:hep-ph/0204202].
P.~Kanti, S.~Lee and K.~A.~Olive,
arXiv:hep-ph/0209036.
\bibitem{casimir}
J.~Garriga, O.~Pujolas and T.~Tanaka,
Nucl.\ Phys.\ B {\bf 605}, 192 (2001)
[arXiv:hep-th/0004109].
R.~Hofmann, P.~Kanti and M.~Pospelov,
Phys.\ Rev.\ D {\bf 63}, 124020 (2001)
[arXiv:hep-ph/0012213].
\bibitem{Charmousis:1999rg}
C.~Charmousis, R.~Gregory and V.~A.~Rubakov,
Phys.\ Rev.\ D {\bf 62}, 067505 (2000)
[arXiv:hep-th/9912160].
\bibitem{Csaki:2000zn}
C.~Csaki, M.~L.~Graesser and G.~D.~Kribs,
Phys.\ Rev.\ D {\bf 63}, 065002 (2001)
[arXiv:hep-th/0008151].
\bibitem{phi2}
S. Deser, Annals of Physics\ {\bf 59}, 248 (1970);
F.~Englert, C.~Truffin and R.~Gastmans, 
Nucl.\ Phys.\ B {\bf 117}, 407 (1976);
J.J.~van~der~Bij, 
Acta Phys. Polon.\ B {\bf 25}, 827 (1994);
R.~Raczka, M.~Pawlowski, 
Found. Phys.\ {\bf 24}, 1305 (1994)  
[arXiv:hep-th/9407137].

\bibitem{Dominici:2002jv}
D.~Dominici, B.~Grzadkowski, J.~F.~Gunion and M.~Toharia,
arXiv:hep-ph/0206192.
\end{thebibliography}
\end{document}